\documentclass[]{raa}            % referee version: for submission
\pdfoutput=1
\usepackage{graphicx,times}             %for PS/EPS graphics inclusion, new
\usepackage{natbib}
\usepackage{amssymb,amsmath}
\bibpunct{(}{)}{;}{a}{}{,}

\newcommand{\figwidth}{0.6\linewidth}

\newcommand{\citepme}{(Xing et al. in preparation)}
\newcommand{\citetme}{Xing et al. (in preparation)}

\usepackage[pagebackref=true]{hyperref}

\begin{document}

  \title{Revisiting The Mass-Size Relation Of Structures In Molecular Clouds
}
%   \subtitle{I. Place Your Subtitle Here}

   \volnopage{Vol.0 (20xx) No.0, 000--000}      %%preserved for Editor. DOn't remove!
   \setcounter{page}{1}          %%starting page, preserved for Editor. DOn't remove!

   \author{Yuchen Xing % Put your Chinese name in "( )" if you like. Note to open line 11 "\usepackage[UTF8]{ctex}"
      \inst{1,2}
   \and Keping Qiu
      \inst{1,2}
   }
%% Here is an example of three authors come from different institutes.
%% For single author or all the authors from an institute, use "\inst{}" only

   \institute{School of Astronomy and Space Science, Nanjing University, Nanjing {\rm 210023}, P. R. China; {\it kpqiu@nju.edu.cn}\\
%% Please give the E-mail address of the author, to whom future correspondence and
%% offprint requests will be sent.
        \and
             Key Laboratory of Modern Astronomy and Astrophysics (Nanjing University), Ministry of Education, Nanjing {\rm 210023}, P.R.China\\
\vs\no
   {\small Received 20xx month day; accepted 20xx month day}}

\abstract{We revisit the mass-size relation of molecular cloud structures based on the column density map of the Cygnus-X molecular cloud complex. We extract 135 column density peaks in Cygnus-X and analyze the column density distributions around these peaks. The averaged column density profiles, $N(R)$, around all the peaks can be well fitted with broken power-laws, which are described by an inner power-law index $n$, outer power-law index $m$, and the radius $R_{\rm TP}$ and column density $N_{\rm TP}$ at the transition point. We then explore the $M-R$ relation with different samples of cloud structures by varying the $N(R)$ parameters and the column density threshold, $N_0$, which determines the boundary of a cloud structure. We find that only when $N_0$ has a wide range of values, the $M - R$ relation may largely probe the density distribution, and the fitted power-law index of the $M-R$ relation is related to the power-law index of $N(R)$. On the contrary, with a constant $N_0$, the $M - R$ relation has no direct connection with the density distribution; in this case, the fitted power-law index of the $M - R$ relation is equal to 2 (when $N_0\ge N_{\rm TP}$ and $n$ has a narrow range of values), larger than 2 (when $N_0\ge N_{\rm TP}$ and $n$ has a wide range of values), or slightly less than 2 (when $N_0< N_{\rm TP}$).
}

   \authorrunning{Y. Xing \& K. Qiu}            %author_head in even pages
   \titlerunning{Revisiting The Mass-Size Relation}  % title_head in odd pages

   \maketitle
%% The author head (on even pages) and the title head (on odd pages) will be
%% automatically extracted from \author{} and \title{}. Whenever the title is too long,
%% you will be asked to supply a shorter one by inserting either \authorrunning{} or
%% \titlerunning{} before \maketitle. Anyway, you can specify your own heads.
%%
%%
%% Note: In the following text body of your manuscript, please note several differences from
%%       other major journals:
%% (1) \subsection{Please Capitalize the First Letter of Each Notional Word in Subsection Title}
%% (2) Please Capitalize the First Letter of Each Notional Word in all tables' captions

%
%________________________________________________ sections below
%

\section{Introduction}

The density distribution reflects the physical state of a molecular cloud thus is important for understanding star formation. However, both volume and column density distributions are difficult to obtain in large quantities directly. Dust extinctions in optical and near-infrared bands can be used to derive $\rm H_2$ distribution at high resolution but cannot probe dense regions (\citealt{Lada1994,Lombardi2001}).
Although dust continuum and molecular lines at millimeter and sub-millimeter wavelengths are free of this problem, they are limited by the low resolution of single-dish radio telescopes and the small dynamic range of interferometers (\citealt{Kellermann2001}). 
Moreover, obtaining the density distributions of a large number of sources across orders of magnitude in density and size is always time-consuming regardless of the observation method used. For decades, the mass-size relation between different structures (hereafter, the $M-R$ relation) has been an important way to explore the density distribution of molecular gas.

An early result of the $M-R$ relation comes from \cite{Larson1981}. Their famous Larson Third law indicated that the density – size relation at $0.1-100\rm\space pc$ is $n(\rm H_2)\propto L^{-1.10}$, corresponding to $M\propto R^{1.9}$. The relation was considered to represent a density distribution of $\rho\propto R^{-1}$, implying that the structures they used in obtaining the $M-R$ relation have approximately the same averaged column density. Since then, there have been a number of observational studies deriving a variety of $M-R$ relations from $M\propto R^{1.4}$ to $M\propto R^{3.0}$, which have been interpreted as $\rho\propto R^{-\alpha}$ distributions with $\alpha=0-1.6$. The $M\propto R^2$ relation is the most commonly seen relation and has been observed in all scales from $10^{-2}\space\rm pc$ to $10^{2}\space\rm pc$ (\citealt{Larson1981,Schneider2004,Lada2020,Mannfors2021}), while the other indexes are mainly observed at $10^{-2}-10^{1}\space\rm pc$ (\citealt{Roman-Duval2010,Urquhart2018,Traficante2018,Massi2019,Lin2019}). 

However, how reliable or accurate the $M-R$ relations are probing the density distributions is still a matter of debate. Observational biases, including the sensitivity limit (\citealt{kegel1989,Schneider2004}) and the column density selection effects for certain tracers (\citealt{Scalo1990, Ballesteros-paredes2002}), as well as the source extraction methodologies (\citealt{kegel1989,Schneider2004,Heyer2009}), can all play a role in the derived $M-R$ relations, and thus affect the inferred density distributions. 
In the 2000s, dust continuum surveys brought new opportunities to understand the $M-R$ relation (\citealt{Enoch2006,Pirogov2007}). The advent of the Herschel observatory (\citealt{Pilbratt2010}) made it possible to map simultaneously extended and compact dust continuum emissions at multi-wavelengths in the far-infrared to sub-millimeter window. Consequently, the column density profiles (hereafter, $N(R)$ profiles) of dense molecular cloud structures can be derived at moderate angular resolutions (\citealt{Arzoumanian2011,Kauffmann2010_MR,Schneider2013}). The $N(R)$ profiles are found to have different indexes at different scales and their corresponding $M(R)$ profiles may be inconsistent with the $M-R$ relation (\citealt{Pirogov2009,Lombardi2010,Kauffmann2010_MR,Beaumont2012}). 
\cite{Lombardi2010}, \cite{Beaumont2012}, and \cite{Ballesteros-Paredes2012} pointed out that measuring the $M-R$ relations based on observations in general implies an effective column density threshold, which in turn would naturally lead to a $M \propto R^2$ relation for typical column density probability distribution functions (N-PDFs), such as log-normal (\citealt{Lombardi2010,Beaumont2012}), power-law (\citealt{Ballesteros-Paredes2012}) or log-normal + power-law (\citealt{Ballesteros-Paredes2012}) N-PDFs. However, there has been no study that links real observational $M(R)$ profiles with $M-R$ relations through mathematical calculations. %thus the differences of $N(R)$ indexes and positions between structures may be ignored. 

In this paper, using the Cygnus-X column density map from \cite{Cao2019}, we obtain $N(R)$ profiles of 135 dense structures at $0.1-10\space \rm pc$. It enables us to derive $M-R$ relations from real density profiles, thus deepening the understanding of the $M-R$ relations, density distributions, and the physical states behind them. We present the obtained $N(R)$ profiles and their parameter distributions in Section~\ref{sec:cyg_range}. In Section~\ref{sec:discuss}, we study effects of the $N(R)$ profiles and column density threshold $N_0$ on the $M-R$ relation. We further discuss the significance of the $N(R)$ profile and the $M-R$ relation from a more realistic perspective in Section~\ref{sec:physic}. The results are summarized in Section~\ref{sec:conclu}.

\section{$N(R)$ profiles of Cygnus-X}\label{sec:cyg_range}

\begin{figure}[h]
    \centering
    \includegraphics[width=\linewidth]{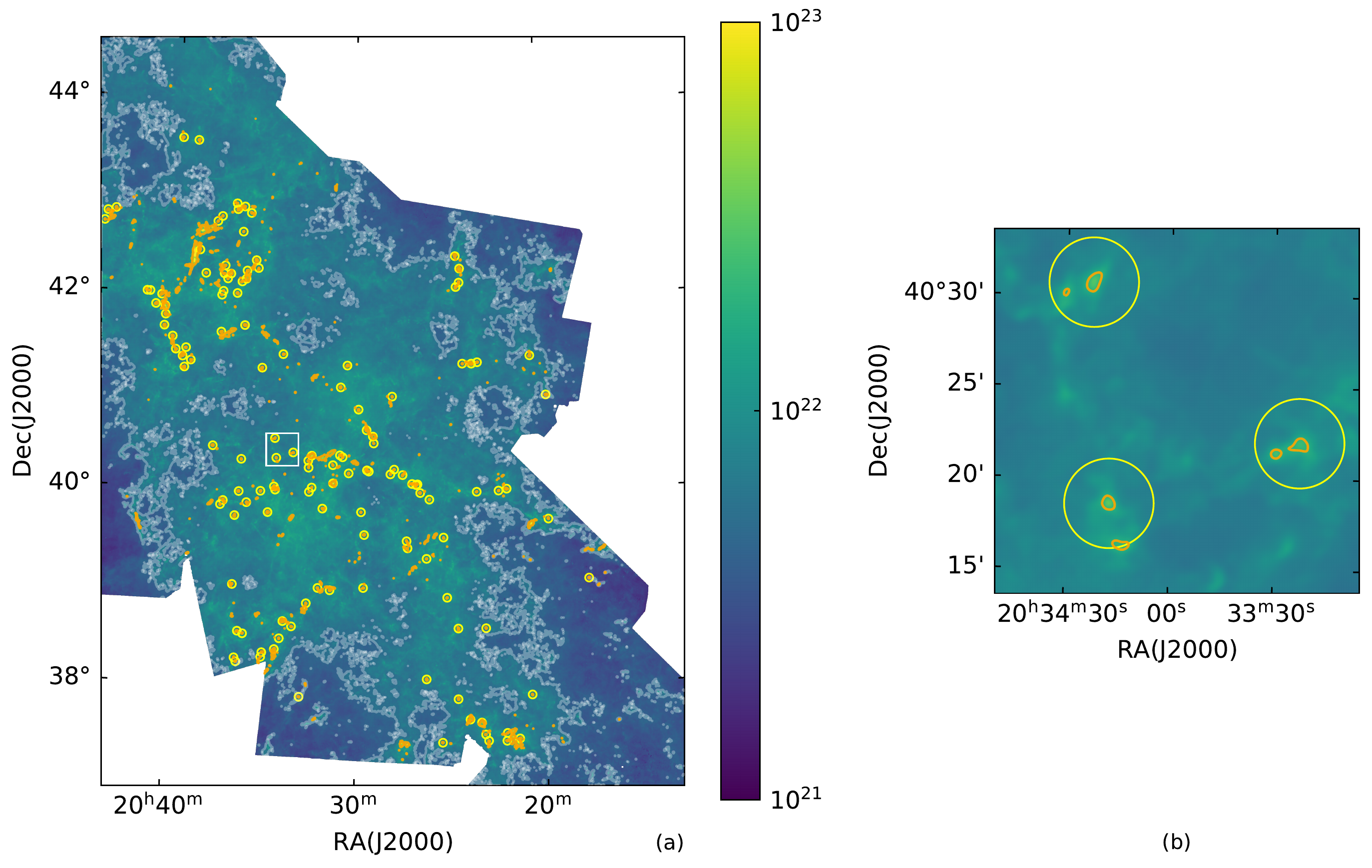}
    \caption{The column density map of Cygnus-X in [$\rm cm^{-2}$]. $a)$ Yellow circles denote the 1 pc radius around the 135 peaks. White contours outline the regions with column densities higher than $5.0\times 10^{21} \rm\space cm^{-2}$. Orange contours outline the regions with column densities higher than $1.86\times 10^{22} \rm\space cm^{-2}$. $b)$ A zoom-in image of the area outlined by the white rectangle in panel $(a)$, to better display the morphology of the column density distribution around the selected density peaks. }
    \label{fig:Nmap_peak}
\end{figure}

\begin{figure}[h]
    \centering
    \includegraphics[width=\figwidth]{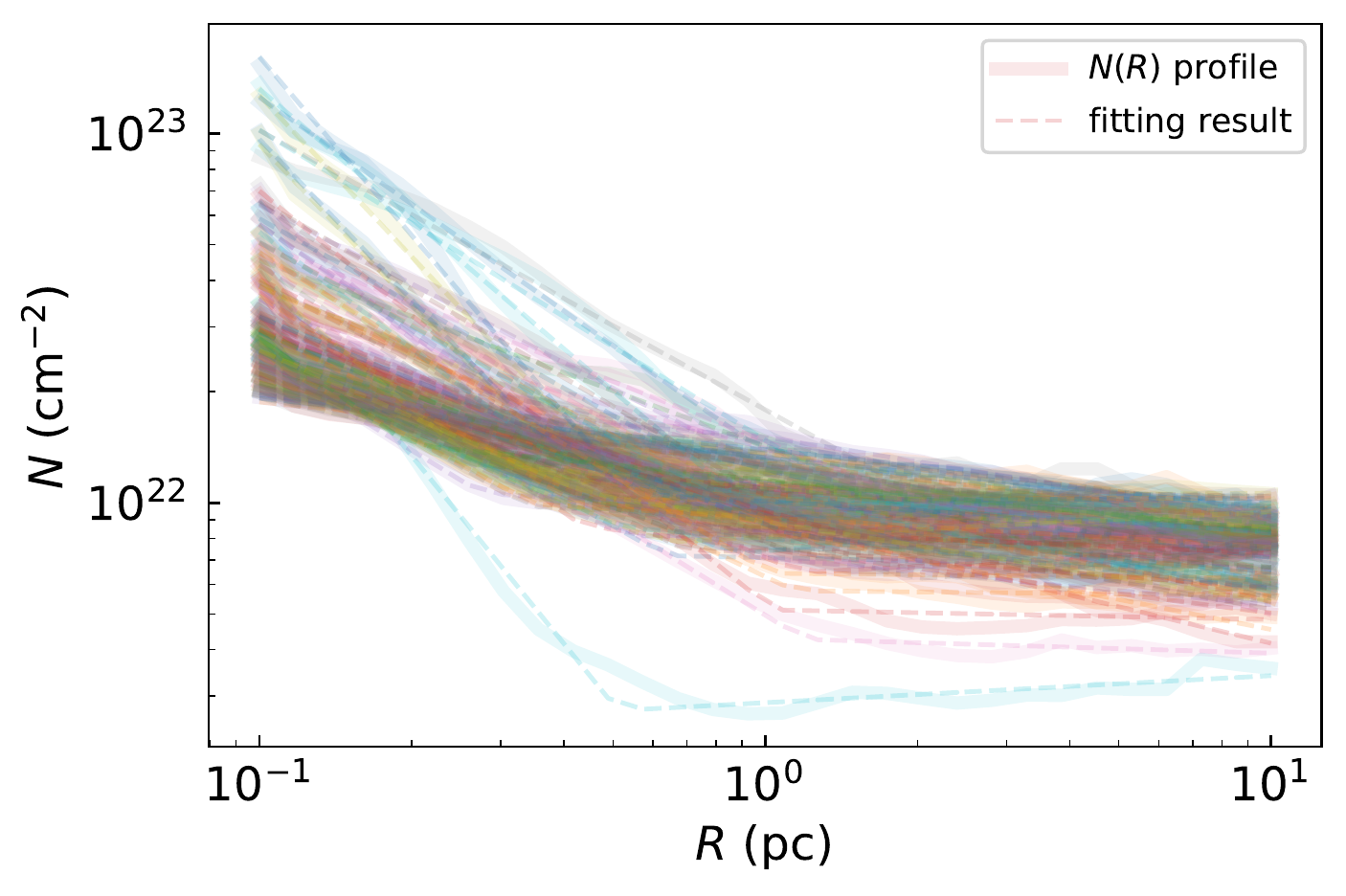}
    \caption{Radial $N(R)$ profiles of the $135$ clumps at $R=0.1-10$ pc. The corresponding broken power-law fittings are shown in dashed lines.}
    \label{fig:radial_N}
\end{figure}

Cygnus-X is one of the most massive giant molecular clouds in our Galaxy (\citealt{Motte2018}), and shows rich star formation activities evidenced by numerous HII regions, OB associations, dense molecular gas clumps and cores (\citealt{Wendker1991,Uyaniker2001,Motte2007,Cao2019,Wang2022}). It is located at a distance of 1.4 kpc from the Sun (\citealt{Rygl2012}). 
Using $getsources$ (\citealt{MenShchikov2012}), \cite{Cao2019} applied SED fittings to the 160, 250, 350, and 500 $\mu \rm m$ dust continuum images from $Herschel$, and obtained the temperature map and the column density map of Cygnus-X. The resolution of the column density map was set by the SED fitting of the smallest spatial scale component using the 160 and 250 um data, and is 18."4 determined by the 250 $\mu$m images, corresponding to 0.1 pc at the distance of 1.4 kpc.  
Using the column density map, \citetme{} obtained the N-PDF of the complex, which shows a log-normal + power-law shape. The turbulence-dominated log-normal component and the gravity-dominated power-law component are delimited by a transitional column density at $1.86\times 10^{22}\space \rm cm^{-2}$ \citepme{}. 
We selected all column density peaks above $1.86\times 10^{22}\space \rm cm^{-2}$ for the extraction of density profiles. To avoid the influence of structures that cannot be described by radial density profiles, we check the morphology of every structure within a density threshold of $1.86\times10^{22}\space\rm cm^{-2}$, and exclude those with aspect ratios larger than 2. In this way, we eventually obtained 135 peaks which are shown in Figure~\ref{fig:Nmap_peak}. We divide the area around each column density peak into 24 sectors with the same angular size (i.e., 15 degrees). For each sector, we calculate the distance of each pixel to the column density peak and average all pixels with the same distance to obtain a sectorized radial $N(R)$ profile. Thus for each column density peak we have 24 $N(R)$ profiles extracted from $10\rm\space pc$ down to the resolution at $0.1\space \rm pc$. We then discard any sectorized profiles that show a column density rise of more than $6.46\times 10^{21}\space\rm cm^{-2}$, which is the peaking column density of the log-normal part in Cygnus-X's N-PDF, with the increasing radius to bypass the contamination from nearby sources. We average the remaining sectorized $N(R)$ profiles to obtain a final $N(R)$ profile for each column density peak. The obtained $N(R)$ profiles are shown in Figure~\ref{fig:radial_N}. In the sections below, to distinguish from the structures used in the $M-R$ relation, we call these 135 structures extending outward from the column density peaks to about $10\space\rm pc$ the 135 Cygnus-X clumps. Note that they are not `clumps' in the usual definition, and have no strict boundaries.

\begin{figure}[h]
    \centering
    \includegraphics[width=\figwidth]{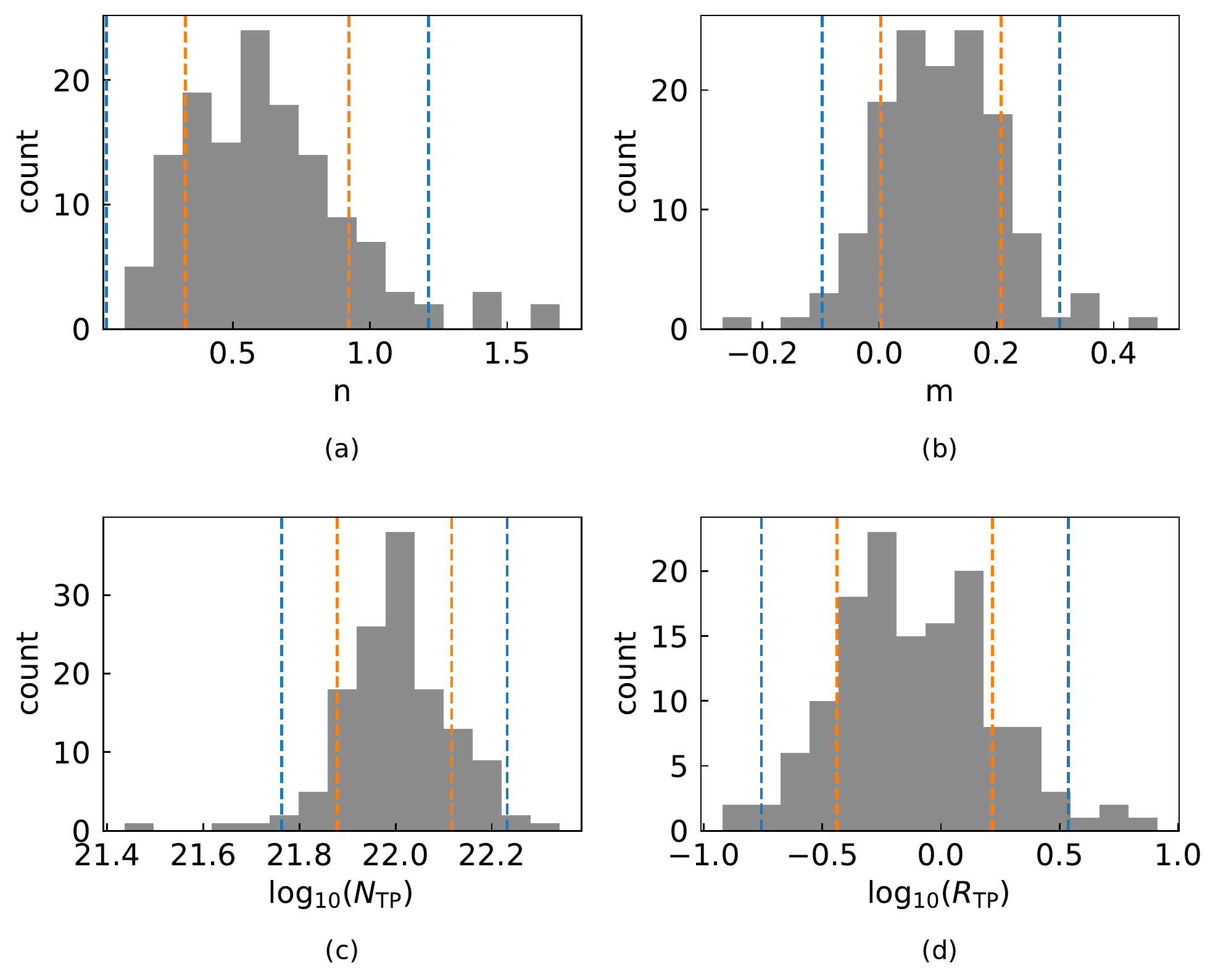}
\caption{Parameters derived by performing broken power-law fittings to the 135 Cygnus-X clumps. Orange and blue dashed lines show the $1\sigma$ ($68\%$) and $2\sigma$ ($95\%$) distribution intervals, respectively. $a)$ The distribution of the inner power-law index $n$. $b)$ The distribution of the outer power-law index $m$. $c)$ The distribution of the transitional column density $N_{\rm TP}$. $d)$ The distribution of $R_{\rm TP}$, the radius at the transitional point.}
\label{fig:para_hist}
\end{figure}

The $N(R)$ profiles are apparently steep in the inner part and flat in the outer part, with the break points roughly around 1 pc (see Figure~\ref{fig:radial_N}). We then fit the $N(R)$ profiles with a broken power-law distribution as described by 
\begin{equation}
    N=
	\begin{cases}
	N_{\rm TP} \left(\frac{R}{R_{\rm TP}}\right)^{-n} & R \le R_{\rm TP} \\
	N_{\rm TP} \left(\frac{R}{R_{\rm TP}}\right)^{-m} & R> R_{\rm TP}
	\end{cases}
	\label{equ:nr}
\end{equation}
where $n$ is the power-law index of the inner region, $m$ the power-law index of the outer region, and $N_{\rm TP}$ and $R_{\rm TP}$ are the transitional column density and radius at the break point, respectively. We obtain good fittings for all the 135 clumps with $R$-squared values all above 0.93. %The $N(R)$ profiles of the 135 clumps meet the broken power-law distribution well
Figure~\ref{fig:para_hist} shows the histograms of the four $N(R)$ parameters of the fitting result.\footnote{Note that all radii we use refer to the full radii instead of half-widths at half maximums which are often used in core statistics.} %Their values reflect the dense core + diffuse cloud nature of Cygnus-X. %Approximately regarding the four parameters to have Gaussian distributions, we obtain the $95\%$ ($2\sigma$) distribution intervals as below.
$95\%$ of the clumps have power-law index $n=0.63\pm 0.59$ at radius $R\le R_{\rm TP}$, corresponding to $M(R)\propto R^{1.37\pm 0.59}$ and $\rho\propto R^{-(1.63\pm 0.59)}$ assuming spherical symmetry. These profiles are close to the free-fall collapse which has $\rho\propto R^{-\alpha}$ with $\alpha=1.5-2.0$, suggesting their gravity-dominated nature. 
20 clumps have $n$ larger than $1$. The largest $n$ goes up to $1.69$, corresponding to a steep density profile with $\alpha=2.69$, which is far beyond the free-fall collapse. At $R>R_{\rm TP}$, the clumps have similar column densities. The outer $N(R)$ index $m$ has a tight distribution, with a $95\%$ distribution interval of $m=0.11\pm 0.20$. It corresponds $M(R)\propto R^{1.89\pm 0.20}$ and $\rho\propto R^{-(1.11\pm 0.20)}$, suggesting the turbulence-dominated nature. %These flat $N(R)$ distributions correspond to the log-normal N-PDF at low column density, representing the turbulence-dominated clouds. 
$R_{\rm TP}$ and $N_{\rm TP}$ have $95\%$ distribution intervals of $R_{\rm TP}=0.78_{-0.60}^{+2.69}\space\rm pc$ and $N_{\rm TP}=1.00^{+0.70}_{-0.41}\times 10^{22}\space\rm cm^{-2}$, respectively. These values, of approximately $R=1\rm\space pc$ and $N=10^{22}\space\rm cm^{-2}$, mark the transition between the two components. %The $N_{\rm TP}$ values are slightly lower than the $1.86\times 10^{22}\rm\space cm^{-2}$ transitional column density of the N-PDF of Cygnus-X, 
%We also obtain that the broken power-law $N(R)$ profiles own column densities of $10^{21.95\pm 0.20}\space\rm cm^{-2}$ at $R>R_{\rm TP}$ and $10^{22.20\pm 0.45}\space\rm cm^{-2}$ at $R\le R_{\rm TP}$ (with 95\% confidence intervals). These column densities are the typical column densities of the two components divided by $R_{\rm TP}$ and $N_{\rm TP}$. 

%In the section below, we use the 135 clumps and artificial clumps to illustrate the form of $M-R$ relations under different circumstances. We note three main differences between the $M-R$ relations obtained from these clumps and the $M-R$ relations in reality. First, 

\section{From $N(R)$ profiles to $M-R$ relations}\label{sec:discuss}

\subsection{Obtaining the $M-R$ relation}

%To understand the relation between the $M-R$ relation and the density profile, we inevitably involve these concepts: the density profile, mass profile, and $M-R$ relation. The density profile, including both volume ($\rho$) and column density ($N$) profiles, describes the density distribution at different radii of a structure. In Section~\ref{sec:cyg_range}, we obtained the $N(R)$ profile of 135 clumps in Cygnus-X. Due to their broken power-law morphology, these $N(R)$ profiles are fully described by the four $N(R)$ parameters. 

By integrating the broken power-law $N(R)$ profile, we can obtain the $M(R)$ profile as 
\begin{equation}
    M=2\pi M_{\rm H_2}\times
	\begin{cases}
	\frac{1}{2-n} \frac{N_{\rm TP}}{R_{\rm TP}^{-n}} R^{2-n} & R \le R_{\rm TP} \\
	\frac{1}{2-n} N_{\rm TP}R_{\rm TP}^{2}+\frac{1}{2-m} \frac{N_{\rm TP}}{R_{\rm TP}^{-m}} (R-R_{\rm TP})^{2-m} & R> R_{\rm TP},
	\end{cases}
	\label{equ:mr}
\end{equation}
where $M_{\rm H_2}=3.32\times 10^{-24}\rm g$ is the hydrogen molecule mass. It is clear that, the $M(R)$ profile has a shape close to broken power-law and is fully described by the four $N(R)$ parameters.
%$M(R)$ profiles of the 135 clumps are shown in gray lines in Figure~\ref{fig:MRindex_cases}. 

The well known $M-R$ relation is obtained by intercepting a group of $M(R)$ profiles with some column density threshold $N_0$. In real observations, $N_0$ is determined by either observational limits, such as the detection limit which is typically a few times the noise level, or by the selection effect of a source extraction algorithm (e.g., \citealt{kegel1989,Scalo1990,Ballesteros-paredes2002}). With the column density threshold $N_0$ determined, the mass $M$ and radius $R$ in a $M-R$ relation are defined as
\begin{equation}
    M=2\pi  \space M_{\rm H_2}\times
	\begin{cases}
	\frac{N_{\rm TP}^{2/n}R_{\rm TP}^2}{2-n}N_0^{1-\frac{2}{n}} &N_0 \ge N_{\rm TP} \\
	\left(\frac{1}{2-n}-\frac{1}{2-m}\right)N_{\rm TP}R_{\rm TP}^2  + \frac{N_{\rm TP}^{2/m}R_{\rm TP}^2}{2-m}N_0^{1-\frac{2}{m}} & N_0 < N_{\rm TP}
	\label{equ:r_n0_0}
	\end{cases}
\end{equation}
and
\begin{equation}
    R=
	\begin{cases}
	R_{\rm TP} \left(\frac{N_0}{N_{\rm TP}}\right)^{-1/n} &N_0 \ge N_{\rm TP} \\
	R_{\rm TP} \left(\frac{N_0}{N_{\rm TP}}\right)^{-1/m}  & N_0 < N_{\rm TP}.
	\end{cases}
	\label{equ:r_n0}
\end{equation}
Thus the mass and radius of a structure in a $M-R$ relation are determined by both the four $N(R)$ parameters and the column density threshold $N_0$. 
Further, combining Equation~\ref{equ:r_n0_0} and Equation~\ref{equ:r_n0}, the $M-R$ relation can be given by
\begin{equation}
    M=2\pi  \space M_{\rm H_2}\times
	\begin{cases}
	\frac{N_0}{2-n}R^2 & N_0 \ge N_{\rm TP} \\
	\left(\frac{1}{2-n}-\frac{1}{2-m}\right)N_{\rm TP}R_{\rm TP}^2  + \frac{N_0}{2-m}R^2 & N_0 < N_{\rm TP}.
	\label{equ:MR_rela}
	\end{cases}
\end{equation}
From Equation~\ref{equ:MR_rela}, the $M-R$ relation at $N_0 \ge N_{\rm TP}$ is apparently a function of $R^2$, and is simplified to $M \propto R^2$ if all the clumps have the same inner power-law index $n$ and are trimmed at a constant $N_0$. Otherwise, the $M-R$ relation deviates from $M \propto R^2$ at a degree depending both on the density profile $N(R)$ and the threshold density $N_0$. We explore the $M-R$ relation in detail in the following subsection.

%when ignoring the influences of the $N(R)$ parameters and $N_0$, the $M-R$ relation naturally show $M\propto R^2$. However, if the sources used for the $M-R$ relation have different $N(R)$ parameters or $N_0$,  shape of the $M-R$ relation is determined by both the four $N(R)$ parameters and the column density threshold $N_0$. Moreover, a $M\propto R^2$ relation will be obtained. 
%, we use the 135 clumps and artificial clumps to study the effects of $N_0$ and $N(R)$ parameters on the $M-R$ relation.
%In the next subsections, by applying different $N_0$ to the 135 $M(R)$ profiles and generating artificial profiles, we obtain $M-R$ relations from the broken power-law $N(R)$ profiles of star-forming molecular clouds and thus discuss their underlying significance.

\subsection{Effects of the $N(R)$ parameters and $N_0$}\label{sec:3_2}

Here we study the effects of the density profile $N(R)$ and the column density threshold $N_0$ on the $M-R$ relation. 
For convenience, we start with $N_0\ge N_{\rm TP}$, and then extend the analysis to the $N_0< N_{\rm TP}$ regime. 

\begin{figure}[h]
    \centering
    \includegraphics[width=\figwidth]{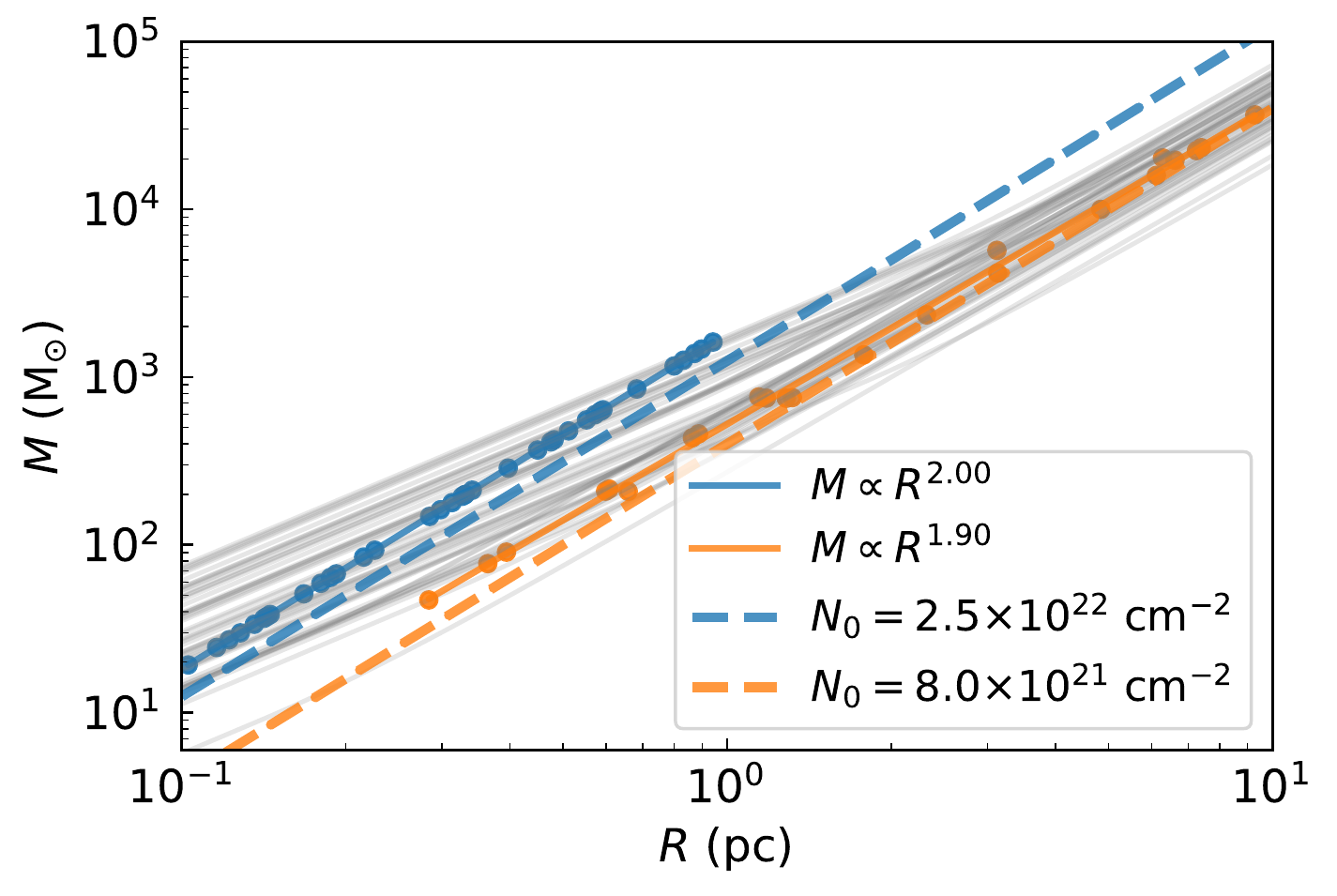}
    \caption{Gray lines show a set of $M(R)$ profiles equivalent to column density profiles that have $n=0.63$, $m=0.11$, and $\sigma_{NR_{\rm TP}}=1$. Blue and orange dots mark the data points obtained by intercepting those $M(R)$ profiles with $N_0 = 2.5 \times 10^{22} \space\rm cm^{-2}$ and $N_0 = 8.0 \times 10^{21} \space\rm cm^{-2}$, respectively. The data points are used to fit the $M-R$ relations (solid lines). Blue and orange dashed lines indicate $M(R)$ profiles corresponding to constant column densities of $2.5 \times 10^{22} \space\rm cm^{-2}$ and $8.0 \times 10^{21} \space\rm cm^{-2}$, respectively. }
    \label{fig:MRcase_all2}
\end{figure}

First, assume that $n$ and $N_0$ are constant. The case of single power-law density profiles with constant $n$ and $N_0$ was analyzed in \cite{Ballesteros-Paredes2012}. It is easy to see from Equation~\ref{equ:MR_rela} that the corresponding $M-R$ relation will show a perfect $M\propto R^2$ shape. Figure~\ref{fig:MRcase_all2} shows an example of this circumstance. All the 50 $M(R)$ profiles in the figure have $n=0.63$ and $m=0.11$. We assume that the $N_{\rm TP}$ and $R_{\rm TP}$ values have Gaussian distributions. We use $\sigma_{NR_{\rm TP}}$ to describe the Gaussian distributions of $N_{\rm TP}$ and $R_{\rm TP}$ with average values of $10^{22}\space\rm cm^{-2}$ and $0.78\space\rm pc$. And the $\mu\pm \sigma$ ranges of $N_{\rm TP}$ and $R_{\rm TP}$ are given by $10^{22.00\pm 0.12\sigma_{NR_{\rm TP}}}\space\rm cm^{-2}$ and $10^{-0.11\pm 0.33\sigma_{NR_{\rm TP}}}\space\rm pc$ (where $\mu$ and $\sigma$ are the mean and standard deviation of a Gaussian distribution). $\sigma_{NR_{\rm TP}}=1$ is used in Figure~\ref{fig:MRcase_all2}. 
We adopt $N_0=2.5\times 10^{22}\space\rm cm^{-2}$ and obtain a sample of structures. We then fit the $M-R$ relation with a power-law model. Only structures within $0.1-10\space\rm pc$ are included in the fitting. The fitting result indeed shows a perfect $M\propto R^2$ relation, which is consistent with the analysis of \cite{Ballesteros-Paredes2012} and the calculations above. For comparison, we lower $N_0$ to $8.0\times10^{21}\space\rm cm^{-2}$, a value slightly below $N_{\rm TP}$, and find that the $M-R$ relation follows $M \propto R^{1.90}$, which is very close to the $M \propto R^2$ relation.

\begin{figure}[h]
    \centering
    \includegraphics[width=\figwidth]{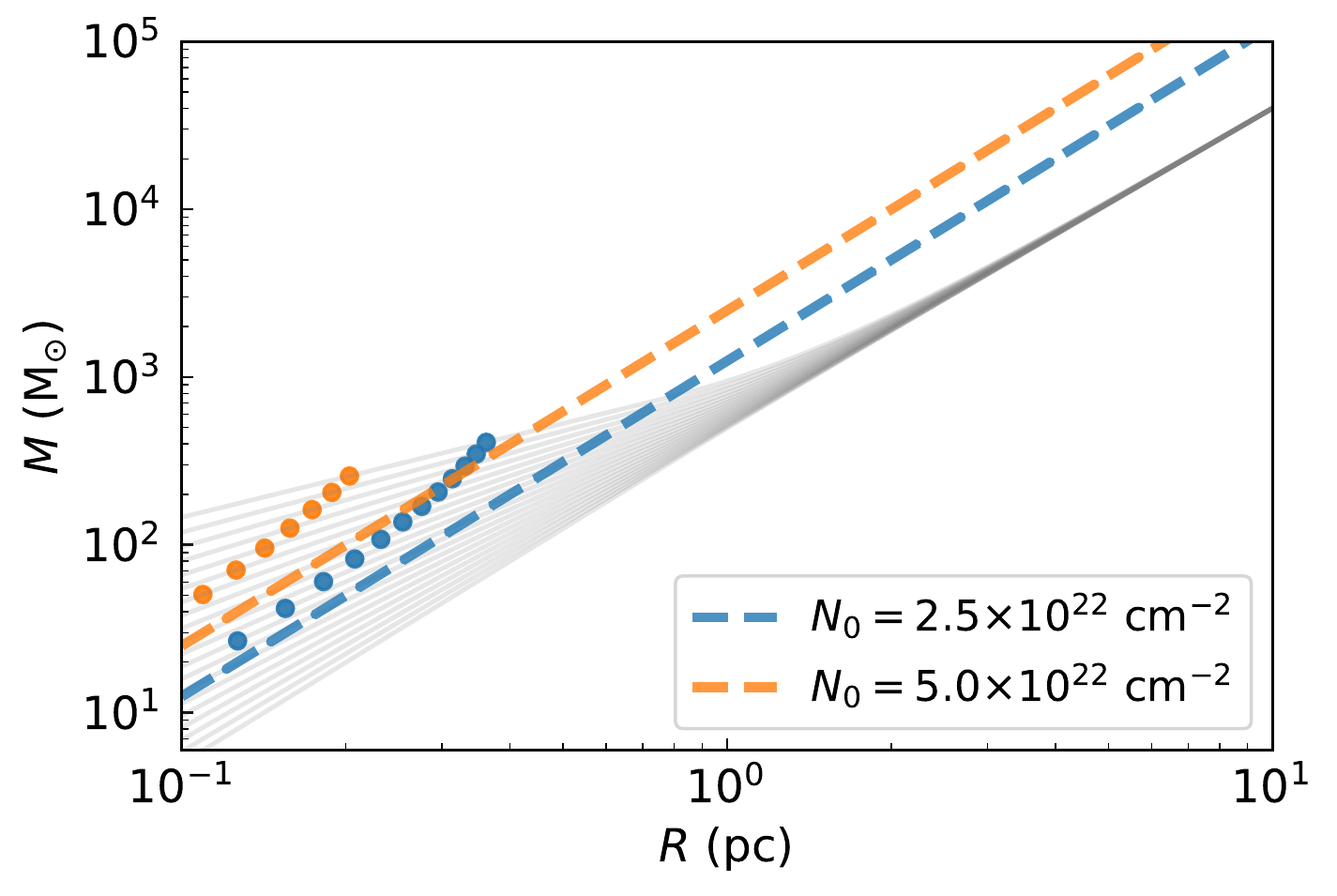}
    \caption{Same as Figure~\ref{fig:MRcase_all2}, but for $m=0.11$, $\sigma_{NR_{\rm TP}}=0$, $n$ evenly distributed from 0 to 1.2, and $N_0 = 2.5 \times 10^{22} \space\rm cm^{-2}$ and $N_0 = 5.0 \times 10^{22} \space\rm cm^{-2}$. }
    \label{fig:MRcase_n}
\end{figure}

We then study the effects of the $N(R)$ index on the $M-R$ relation, i.e., effects of $n$ when $N_0\ge N_{\rm TP}$. We generate 20 density profiles with $m=0.11$ and $\sigma_{NR_{\rm TP}}=0$. Their $n$ values are evenly distributed in the range of $0-1.2$. In Figure~\ref{fig:MRcase_n}, these density profiles correspond to $M(R)$ profiles that are almost the same at larger radii and have different indexes at smaller radii. $M(R)$ profiles with larger $n$ have shallower indexes at $R\le R_{\rm TP}$. 
As in Equation~\ref{equ:MR_rela}, for any $n$ between 0 and 2 (Note that $n$ cannot be equal to 2 in order to obtain a finite mass. $n$ with a value outside the range of $0-2$ is impractical for it will either correspond to a $N(R)$ profile that is denser on the larger radii, or have a density profile steeper than $\rho\propto R^{-3}$), $M$ always increases with $n$. When adopting $N_0=2.5\times 10^{22}\space\rm cm^{-2}$, from bottom to top, the $n$ value increases and the data points deviate more from the $M\propto R^2$ line. The $M-R$ relation is apparently steeper than $M\propto R^2$ and bends upward at the high mass end. While for the case with $N_0=5.0\times 10^{22}\space\rm cm^{-2}$, only 7 structures with the largest $n$ fall in $0.1-10\space\rm pc$. It makes the $n$ range of the used data smaller, so the bending is less obvious. But the $M-R$ relation is still steeper than $M\propto R^2$. %The curve is less obvious in the case with $N_0=5.0\times 10^{22}\space\rm cm^{-2}$, for only the 7 structures with the largest $n$ fall in $0.1-10\space\rm pc$ thus the actual $n$ range of the used data is smaller. But the $M-R$ relation is still steeper than $M\propto R^2$. 

\begin{figure}[h]
    \centering
    \includegraphics[width=\figwidth]{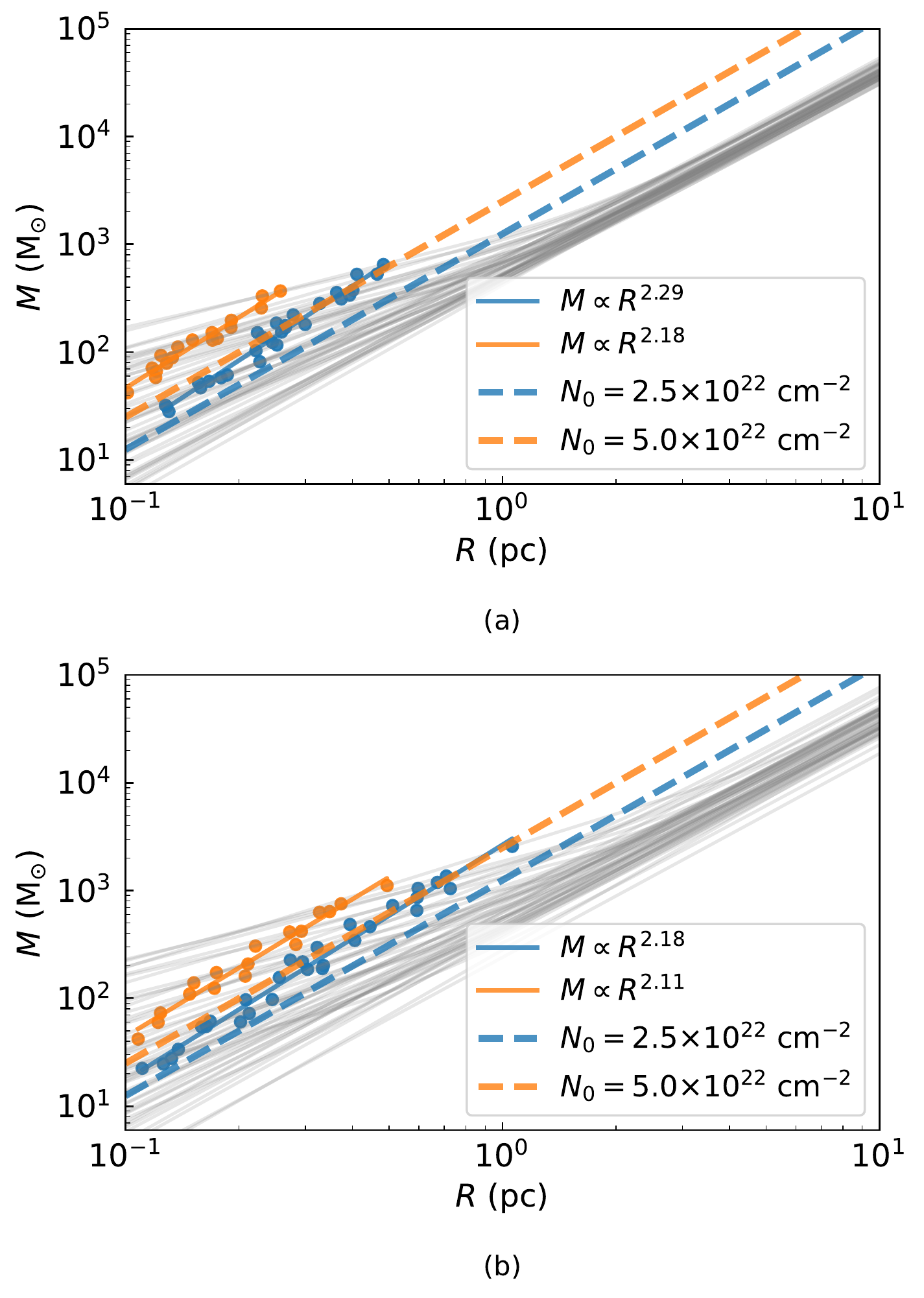}
    \caption{Same as Figure~\ref{fig:MRcase_n}, but for $\sigma_{NR_{\rm TP}}=0.5$ in panel $(a)$ and $\sigma_{NR_{\rm TP}}=1.0$ in panel $(b)$. }
    \label{fig:MRcase_NRmax}
\end{figure}

In Figure~\ref{fig:MRcase_NRmax} we show the effect of $N_{\rm TP}$ and $R_{\rm TP}$ on the $M-R$ relation. As in the $N_0>N_{\rm TP}$ part of Equation~\ref{equ:r_n0_0} and Equation~\ref{equ:r_n0}, both $M$ and $R^2$ are proportional to $N_{\rm TP}^{2/n} R_{\rm TP}^2$. Therefore, %$N_{\rm TP}$ and $R_{\rm TP}$ act as scaling coefficients in the $M-R$ plot.
$N_{\rm TP}$ or $R_{\rm TP}$ with wider ranges will make the data points distribute within a larger range along the $M\propto R^2$ line. It will eventually make the $M-R$ index approaches $2$, regardless of the original value of the $M-R$ index. 
The density profiles in panel $(a)$ of Figure~\ref{fig:MRcase_NRmax} have $n$ and $m$ generated in the same way as those in Figure~\ref{fig:MRcase_n}, but their $N_{\rm TP}$ and $R_{\rm TP}$ follow $\sigma_{NR_{\rm TP}}=0.5$. Adopting $N_0 = 2.5\times 10^{22}\space\rm cm^{-2}$ and $5.0\times 10^{22}\space\rm cm^{-2}$, the $M-R$ relations are found to be closer to $M\propto R^2$ compared to their counterparts in Figure~\ref{fig:MRcase_n}. With $N_{\rm TP}$ and $R_{\rm TP}$ varying over a large range, the bendings caused by $n$ are washed out. We then fit the two $M-R$ relations with power-law models and the results are $M\propto R^{2.29}$ and $M\propto R^{2.18}$. 
We further increase the Gaussian distributions of $N_{\rm TP}$ and $R_{\rm TP}$ to $\sigma_{NR_{\rm TP}}=1$ in panel $(b)$ of Figure~\ref{fig:MRcase_NRmax}. The fitting results of the $M-R$ relation are $M\propto R^{2.18}$ and $M\propto R^{2.11}$, which are closer to $M\propto R^2$ compared to those in panel $(a)$. 

\begin{figure}[h]
    \centering
    \includegraphics[width=\figwidth]{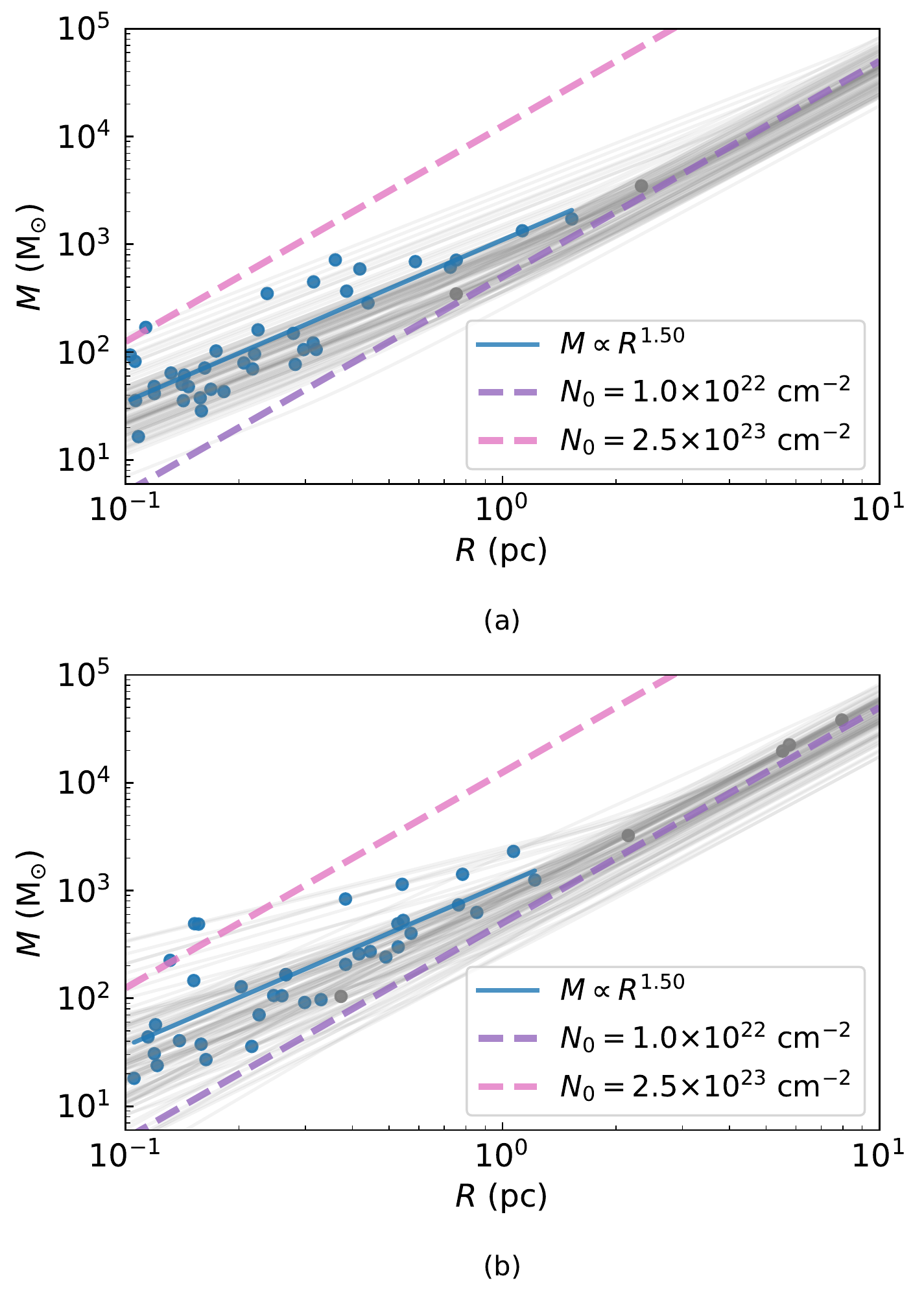}
    \caption{$a)$: Same as Figure~\ref{fig:MRcase_all2}, but for $N_0$ ranging from $1.0 \times 10^{22} \space\rm cm^{-2}$ to $2.5 \times 10^{23} \space\rm cm^{-2}$. The $M(R)$ profiles corresponding to the upper and lower limits of $N_0$ are shown in pink and purple dashed lines, respectively. Blue and gray points correspond to structures with $N_0\ge N_{\rm TP}$ and $N_0<N_{\rm TP}$, respectively. The fitting result of the blue points is shown in the solid blue line. $b)$: same as panel $(a)$, but for $n$ having a Gaussian distribution with $\mu \pm \sigma=0.63\pm 0.30$, where $\mu$ and $\sigma$ are the mean and standard deviation of the Gaussian distribution.}
    \label{fig:MRcase_N0}
\end{figure}

Figure~\ref{fig:MRcase_N0} shows the effects of $N_0$ on the $M-R$ relation. In panel $(a)$ of Figure~\ref{fig:MRcase_N0}, we adopt $n=0.63$, $m=0.11$, and  $\sigma_{NR_{\rm TP}}=1$. Note that the corresponding $M-R$ relation will inevitably be affected by the wide $N_{\rm TP}$ and $R_{\rm TP}$ ranges as discussed above. The $n$ value corresponds to $M(R)\propto R^{1.37}$ at $N\ge N_{\rm TP}$. 
According to Equation~\ref{equ:MR_rela}, $\frac{M}{R^2}$ is proportional to $N_0$, meaning that the variation in $N_0$ would wash out the $M\propto R^2$ relation. In addition, considering an extreme that the clumps under investigation all have an identical density profile, differing $N_0$ is to catch structures falling on different positions along the $M(R)$ profile, and the $M-R$ relation would have a shape the same as that of the $M(R)$ profile provided the sample is large enough and $N_0$ is randomly drawn from a wide range. We adopt $N_0=(0.1-2.5)\times 10^{23}\space \rm cm^{-2}$. The $N_0$ range covers the column densities of most $M(R)$ profiles at $N \ge N_{\rm TP}$, but it also includes some $M(R)$ profiles at $N < N_{\rm TP}$. For simplicity, we only use data points with $N\ge N_{\rm TP}$ for fitting and obtain $M\propto R^{1.50}$. This can be understood as a combined effect of varying $N_{\rm TP}$, $R_{\rm TP}$, and the large range of $N_0$, with the former having a tendency of $M \propto R^2$ and the latter getting the $M - R$ relation close to the $M(R)$ profile. In panel $(b)$ of Figure~\ref{fig:MRcase_N0}, we change $n$ to have a Gaussian distribution with the $\mu \pm \sigma=0.63\pm 0.30$. The obtained $M-R$ index does not change since the averaged value of $n$ is still $0.63$. It is also noticeable that allowing $N_0$ to vary within a range will induce significant  scatter of the data points in the $M-R$ plot. 

Let's then consider the $N<N_{\rm TP}$ part of the $M-R$ relation. Firstly, the $\frac{N_0}{2-m}R^2$ term of the $N<N_{\rm TP}$ part is similar to the $M-R$ relation at $N\ge N_{\rm TP}$ (see Equation~\ref{equ:MR_rela}), thus the $N<N_{\rm TP}$ part has all the effects discussed above. Aside from these, the $N<N_{\rm TP}$ part has an additional $\left(\frac{1}{2-n}-\frac{1}{2-m}\right)N_{\rm TP}R_{\rm TP}^2$ term. It represents the additional mass introduced by the $N\ge N_{\rm TP}$ part and is independent of $N_0$. 
Considering that $n$ is larger than $m$, this term will increase the obtained mass. 
As in Equation~\ref{equ:mr}, for any structure trimmed by a threshold with $N_0<N_{\rm TP}$, the smaller its radius, the larger the proportion of this term in the total mass. Therefore, the existence of this term will shift the left end of the $M-R$ relation more upward, and thus flatten the $M - R$ relation (e.g., $M\propto R^{1.90}$ in the case shown in orange in Figure~\ref{fig:MRcase_all2}). 
%Taking the orange points in Figure~\ref{fig:MRcase_all2} as an example, the data points on the left deviate more from $N_0$ than the rightmost data point at $10\rm\space pc$. 
%From the perspective of the $M-R$ relation, the present of the term shifts its left end upward and leads to an index that is slightly lower than 2 ($M\propto R^{1.90}$ in the case in Figure~\ref{fig:MRcase_all2}). 

Now we can summarize the effects of the $N(R)$ parameters and the column density threshold $N_0$ on the $M-R$ relation as follows: (1) constant $N(R)$ profile power-law index and constant $N_0$ give rise to a $M\propto R^2$ tendency; (2) the $N(R)$ power-law index with a wide range steepens the $M-R$ relation; %and in some cases, bends the relation upward at the high mass end, 
(3) $N_{\rm TP}$ and $R_{\rm TP}$ with wide ranges to some extent weaken the steepening effect due to the variation of the $N(R)$ power-law index; (4) $N_0$ with a wide range tend to wash out the $M\propto R^2$ relation and get the $M-R$ relation approaching the averaged density profile of the sample sources; (5) at $N<N_{\rm TP}$, the fact that $n$ is larger than $m$ makes the $M-R$ index lower than 2.

\subsection{The $M-R$ relation of the 135 Cygnus-X clumps}\label{sec:MR_N0}

\begin{figure*}[t]
    \centering
    \includegraphics[width=\linewidth]{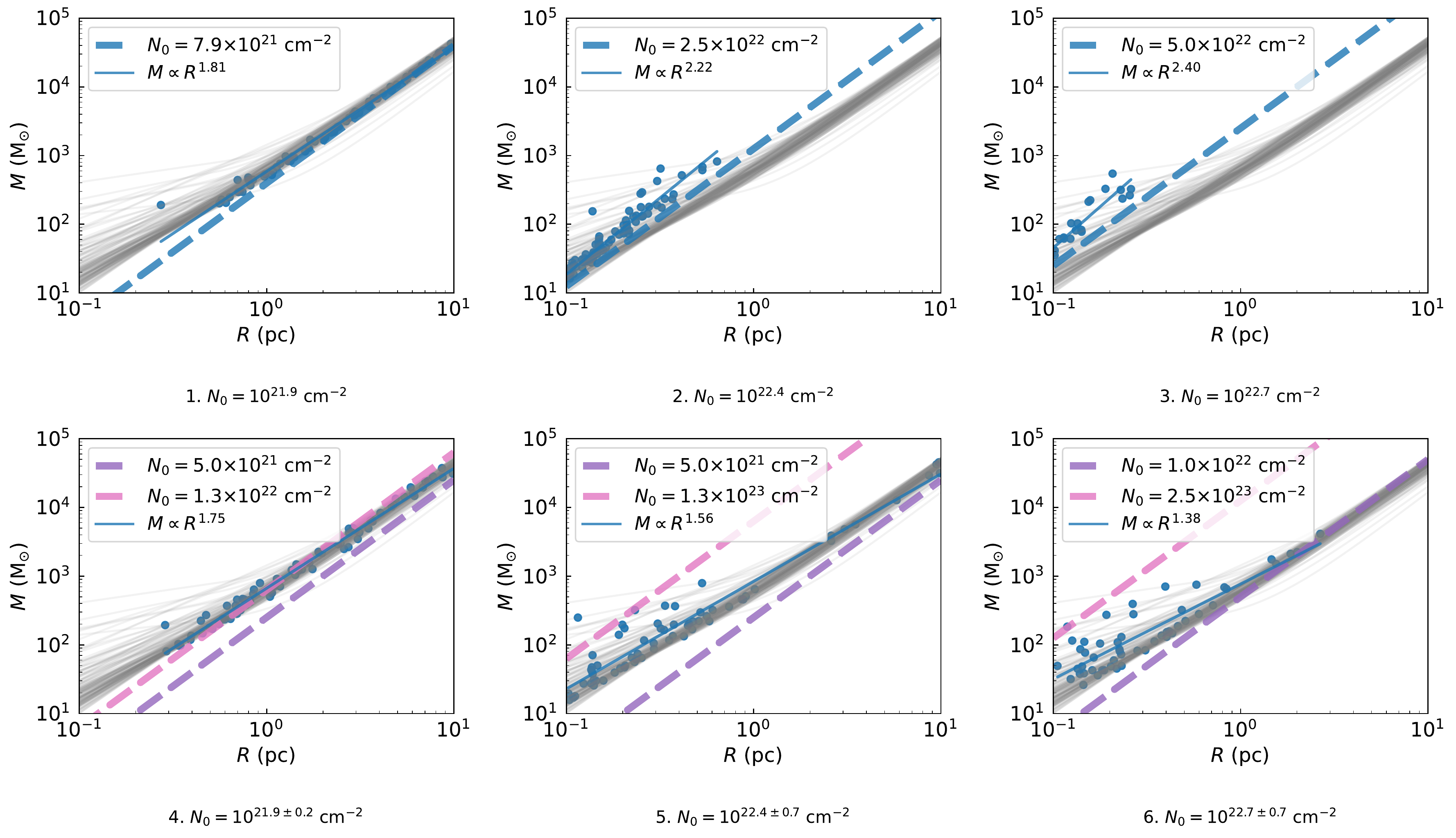}
\caption{$M-R$ relations derived from the $M(R)$ profiles of the 135 Cygnus-X clumps, which are intercepted with different choices of $N_0$ as indicated below each panel. Other symbols are the same as those shown in Figure~\ref{fig:MRcase_all2}-\ref{fig:MRcase_N0}. }
\label{fig:MRindex_cases}
\end{figure*}

Using the 135 Cygnus-X clumps, we look into the $M-R$ relations from a more realistic perspective. The $N(R)$ parameters of the 135 clumps are described in Section~\ref{sec:cyg_range}. 
By fixing $N_0$ at a certain value, or allowing it to vary within a range, we obtain six samples of the clumps. We then fit the $M-R$ relations with power-law models. Only structures within $0.1-10\space\rm pc$ are included in the fitting. The results are shown in Figure~\ref{fig:MRindex_cases}. 

Cases 1-3 have their $N_0$ fixed at a certain value. We adopt $N_0 = 7.9\times10^{21}\space \rm cm^{-2}$ in Case 1. This $N_0$ is lower than $N_{\rm TP}$ of most clumps, and the obtained structures mainly fall at $1 - 10\space\rm pc$. The $M-R$ relation is similar to that shown in orange in Figure~\ref{fig:MRcase_all2}: the constant $N_0$ favors a $M\propto R^2$ relation, while $N_0 < N_{\rm TP}$ makes the $M-R$ index slightly lower. The difference of $m$ between the 135 clumps increases the $M-R$ index, but the increase is negligible since $m$ has a narrow distribution. These all finally lead to a $M\propto R^{1.81}$ relation. In Case 2 and 3, We adopt $N_0 = 2.5\times10^{22}\space \rm cm^{-2}$ and $N_0 = 5.0\times10^{22}\space \rm cm^{-2}$. These $N_0$ are higher than all of the $N_{\rm TP}$ values of the 135 clumps, and the obtained structures all fall at $0.1 - 1\space\rm pc$. The $M-R$ relations are similar to those in Figure~\ref{fig:MRcase_n} and Figure~\ref{fig:MRcase_NRmax}: on top of the $M\propto R^2$ trend contributed by the constant $N_0$, the difference between $n$ makes the $M-R$ relations steeper, leading to $M \propto R^{2.22}$ and $M \propto R^{2.40}$.

In Cases 4-6, $N_0$ of each structure is randomly generated within a range in logarithmic space. Most structures obtained in Case 4 have radii falling in the range of $1-10\rm\space pc$. The $N_0$ range of $10^{21.9\pm 0.2}\space \rm cm^{-2}$ is wide enough to cover the column densities of most clumps below $N_{\rm TP}$. With the wide $N_0$ range, the $M-R$ relation can potentially probe the density profiles in the $N<N_{\rm TP}$ regime. However, there are still some structures with sizes of $0.1-1\rm\space pc$, at which the $M(R)$ profiles have their mean as $M(R)\propto R^{1.37}$. These structures make the obtained $M-R$ relation slightly shallower than the mean $M(R)$ profile of $M(R)\propto R^{1.89}$ for $N<N_{\rm TP}$, and the fitting result is $M\propto R^{1.75}$. In Case 5 we use $N_0 = 10^{22.4\pm 0.7}\space \rm cm^{-2}$. The $N_0$ range covers the column densities of most clumps at $0.1-10\rm\space pc$. In this case, both the $0.1-1\rm\space pc$ and $1-10\space\rm pc$ parts contain a considerable number of structures, and the $M-R$ index at $1.56$ is also in the middle of the mean $M(R)$ profile indexes of the two parts. In Case 6 we adopt $N_0 = 10^{22.7\pm 0.7}\space \rm cm^{-2}$. The $N_0$ range covers the column densities of most clumps at $0.1-1\rm\space pc$. Most of the obtained structures have sizes of $0.1-1\space\rm pc$, and the $M\propto R^{1.38}$ relation is very close to the mean $M(R)$ profile for $N\ge N_{\rm TP}$. 

\begin{figure}[h]
    \centering
    \includegraphics[width=\linewidth]{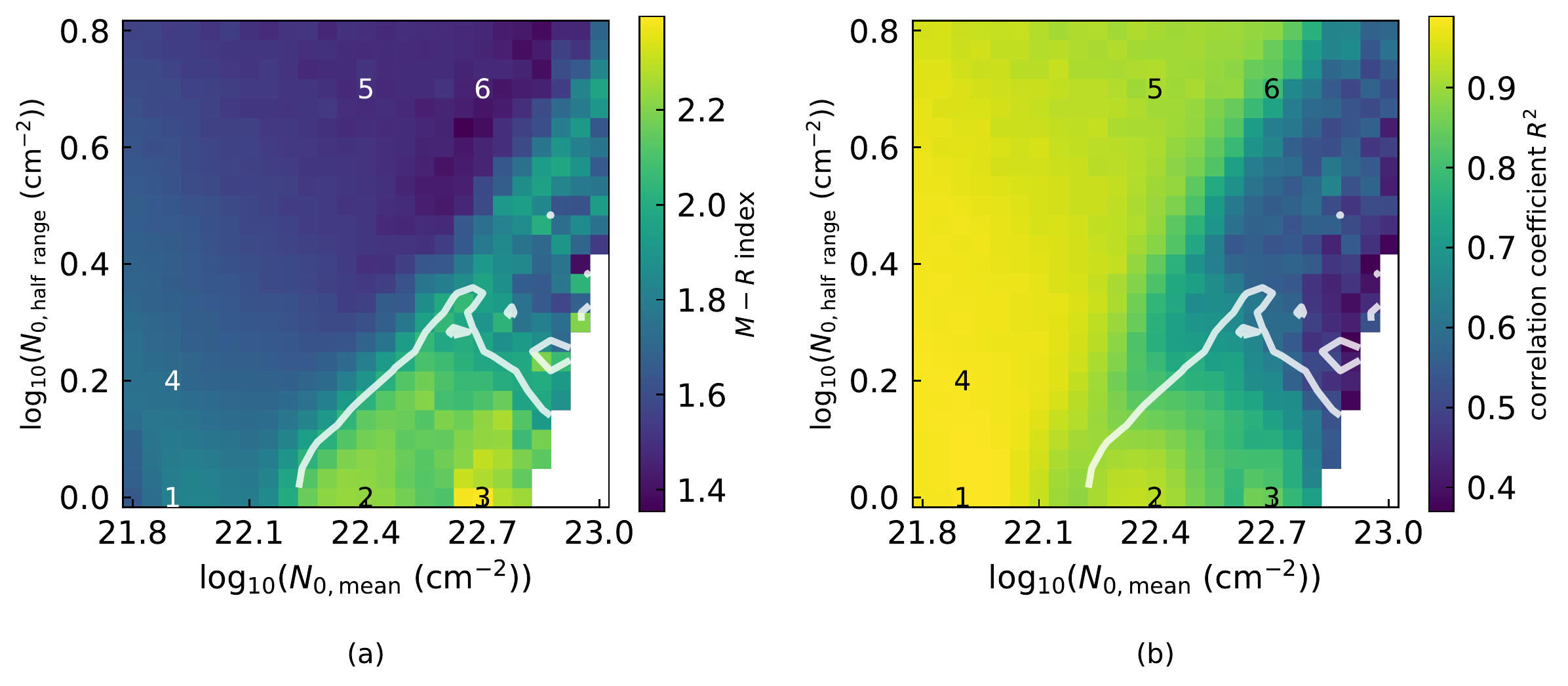}
    \caption{The $M-R$ indexes (left) and the $M-R$ fitting results' correlation coefficients (right) of the 135 clumps. The column density threshold $N_0$ is defined by ${\rm log}_{10}(N_0)={\rm log}_{10}(N_{0,\rm mean})\pm {\rm log}_{10}(N_{0,\rm half\space range})$. The numbers correspond to the cases in Figure~\ref{fig:MRindex_cases}. White contours show $M-R$ indexes of 2. }
%The vertical dashed line shows the mean of $N_{\rm TP}$ at $1.0\times 10^{22}\space\rm cm^{-2}$.
\label{fig:MRindex_cyg}
\end{figure}

We further obtain the $M-R$ indexes in Figure~\ref{fig:MRindex_cyg}. By simply varying the mean and range of $N_0$, $M-R$ relations with indexes from $1.4$ to $2.4$ are obtained from the 135 clumps. 
When $N_0$ is fixed at a certain value, we obtain $M-R$ relations with indexes of $1.8-2.4$. The $M-R$ relations with $N_0<10^{22.1}\space\rm cm^{-2}$ have indexes around $1.8-1.9$. %As in the analysis above, it is the result of both the $M\propto R^2$ tendency of all $M-R$ relations and the fact that $n$ is larger than $m$. 
At higher $N_0$, we obtain $M-R$ relations with indexes of $2-2.4$. %These results are caused by the difference of $n$ between clumps. Due to the limited number of sources, the shape of the $M-R$ relation is easily affected by the positions of individual structures, thus the $M-R$ index fluctuates with the increase of $N_0$. 
When $N_0$ has a sufficiently wide range, the $M-R$ index is always smaller than 2 and decreases with the increase of $N_0$. These indexes manifest the density profiles at the corresponding scales. The decreasing trend comes from the difference between the mean shapes of the two parts of the $M(R)$ profiles, i.e., $M(R)\propto R^{1.89}$ for $R>R_{\rm TP}$ and $M(R)\propto R^{1.37}$ for $R\le R_{\rm TP}$. %With the increase of $N_{0,\rm mean}$, the scale reflected by the $M-R$ relation changes from $R>R_{\rm TP}$ to $R\le R_{\rm TP}$, thus the $M-R$ index decreases. 
When $N_0$ varies within a range but the range is smaller, the $M-R$ index will be between the two cases where $N_0$ is fixed and has a wide range. %and the significance of the $M-R$ relation will also be the combination of the two cases.

\section{Discussion} \label{sec:physic}

%\begin{table*}[t]
    %\tabcolsep 18pt
%    \footnotesize
%    \caption{Physical meaning of the $M-R$ relation with different $N_0$.}
%    \label{tab:mrsummary}
%    \centering
%     \begin{tabular*}{0.9\textwidth}{c||c|c|c}
%        Wide $N_0$ & $\overline{\alpha}=3-k$; & $\overline{\alpha}=3-k$; & \\
%        & unable to obtain the distribution of $\alpha$. & unable to obtain the distribution of $\alpha$. &\\
%        \hline
%        Constant $N_0$ & & $\alpha$ has a tight distribution, & $\alpha$ has a wide distribution,  \\
%        & & unable to obtain the value of $\alpha$. & unable to obtain the value of $\alpha$. \\
%        \hline\hline
%        & $k<2$ & $k=2$ & $k>2$
%    \end{tabular*}
    %\caption{$k$ is the index of the $M-R$ relation. $\alpha$ and $\overline{\alpha}$ are the index and the mean index of the density profile $\rho\propto R^{-\alpha}$. }
%    \begin{tablenotes}
%    \item[1)] $k$ is the index of the $M-R$ relation
%    \item[2)] $\alpha$ and $\overline{\alpha}$ are the index and the mean index of the density profile $\rho\propto R^{-\alpha}$
%    \end{tablenotes}
%\end{table*}

In Section~\ref{sec:cyg_range}, we obtained the column density profiles of 135 clumps in Cygnus-X. Their main features are: 1) the profiles all show broken power-law shapes at $0.1-10\space\rm pc$, 2) the parameters of each clump's profile are different, 3) the transition points of the profiles are around $0.8\rm\space pc$ and $10^{22}\space\rm cm^{-2}$, suggesting density profiles of $\rho\propto R^{-1.63}$ in the inner part and $\rho\propto R^{-1.11}$ in the outer part. 
These two parts of density profiles are consistent with the circumstances of free-fall collapse ($\rho\propto R^{-\alpha}$, with $\alpha=1.5-2.0$) and turbulence dominated nature ($\alpha\approx1.0$), respectively. They are also comparable to the log-normal + power-law N-PDF of Cygnus-X. The N-PDF of Cygnus-X has its power-law index at $2.33$ \citepme{}, corresponding to a density profile of $\rho\propto R^{-1.86}$. The transitional column density between the log-normal and power-law parts is at $1.86\times 10^{22}\space\rm cm^{-2}$. These values are slightly different from the parameters of our density profiles, which is because the power-law index of a N-PDF is more affected by the densest sources, and the transitional column density is related to the proportion of high and low density components. These all suggest that the broken power-law column density profiles imply a gravity-dominated dense core + turbulence-dominated diffuse cloud, and the transition point at $0.8\rm\space pc$ and $10^{22}\space\rm cm^{-2}$ acts as the division of the two components. 

In Section~\ref{sec:discuss} we show how the $N(R)$ profiles and $N_0$ affect the shape of the $M-R$ relation. From the observational point of view, %with the $N(R)$ parameters determined by the observational target, the significance of the $M-R$ relation depends on the column density threshold $N_0$. 
$N_0$ is often determined by the detection limit or set by a threshold in some source extraction method (\citealt{kegel1989,Scalo1990,Ballesteros-paredes2002}), and it is likely to be constant.  
%when the $N_0$ is determined by sensitivity (\citealt{Ballesteros-paredes2002,Roman-Duval2010,Kainulainen2011}), the tracer's lower column density limit (\citealt{scalo1990,Ballesteros-paredes2002}), or some source extraction methods (\citealt{Roman-Duval2010,Kainulainen2011}), it will likely be constant. 
In such a situation the $M-R$ relation may show the well-known $M\propto R^2$ scaling law (\citealt{Lombardi2010,Ballesteros-Paredes2012}), provided $N_0 \ge N_{\rm TP}$ and the power-law index $n$ is nearly constant; the $M-R$ power-law indexes may be slightly less than 2 if $N_0$ is lower than $N_{\rm TP}$ (Figure~\ref{fig:MRindex_cases}.1); the $M-R$ relation may also appear to be steeper than $M\propto R^2$, as seen in some other studies (\citealt{Roman-Duval2010,Kainulainen2011}), when $N_0 \ge N_{\rm TP}$ and the power-law index $n$ varies from source to source. 

When many molecular cloud structures are included in an analysis (\citealt{Larson1981,Urquhart2014,Urquhart2018}), $N_0$ is likely to have a wide distribution, if the cloud structures under investigation were obtained from different observations or extracted from highly varying backgrounds. In these cases, the derived $M-R$ relations may to some extent manifest the density profiles of the cloud structures. But caution should be taken in converting the observed $M-R$ relation to a density profile if the observations significantly suffered from short dynamical ranges (e.g., \citealt{Scalo1990,Ballesteros-paredes2002,Schneider2004}). For molecular cloud structures in Cygnus-X, a $M\propto R^{1.9}$ relation is expected to be obtained with an observational study capable of trimming the structures at large and varying radii (e.g., at $R>1\rm\space pc$ and $N<10^{22}\rm\space cm^{-2}$), as a consequence of a tight distribution of the power-law indexes for the density profiles in the outer parts (i.e., $\rho\propto R^{-1.1}$, see Section~\ref{sec:cyg_range}). Shallower relations can be found at smaller scales and higher densities. They correspond to $\rho\propto R^{-\alpha}$ density distributions with $\alpha$ clearly larger than 1.

%For $N_0$ determined in other ways, e.g. $N_0$ determined by algorithms, it may be difficult to tell whether its distribution is wide or narrow directly.
Applying different source extraction or identification algorithms to the same molecular cloud, different structures (\citealt{Schneider2004,Lichong2020}) and $M-R$ relations (\citealt{Schneider2004}) can be obtained. Without knowing the impact on $N_0$ of the source extraction process, it is difficult to make convincing interpretations of the $M-R$ relations. For example, $\rho\propto R^{-1}$ profiles and a constant $N_0$ can both result in $M\propto R^2$ relations. %but they correspond to $N_0$ with narrow and wide distributions respectively. 
Stronger line-of-sight contamination for larger cloud structures (\citealt{Ballesteros-Paredes2019}), an approximately constant volume density for all the cloud structures under investigation (\citealt{Lada2008,Li2020}), and variation from source to source in the power-law index of the $N(R)$ profiles can all lead to $M-R$ relations steeper than $M\propto R^2$, but only when $N_0$ is constant or falls in a narrow range can a steep $M-R$ relation come from the difference in the $N(R)$ index. Therefore, before interpreting an $M-R$ relation, one is suggested to carefully check how the cloud structures in the sample are derived and then to determine if a constant $N_0$, or instead a varying $N_0$, is implicitly being used; only in the latter case, the observed $M-R$ relation could be useful in constraining the averaged density profile of the cloud structures under investigation. From another perspective, an observational experiment optimized for converting a $M-R$ relation to a density profile would require high resolution and high sensitivity to reasonably resolve each source and allow an estimate of the source flux and size free of sensitivity limitation (e.g., estimation based on the peak intensity and FWHM size by 2D Gaussian fitting to the source brightness distribution). This way one equivalently has $N_0$ varying from source to source. High resolution also helps to minimize potential line-of-sight contamination, while high sensitivity observations of optically thin tracers are desirable to increase the dynamical range. 

%Whether $M-R$ relations demonstrate virial equilibrium has been a concern.
Can $M-R$ relations help to determine whether the cloud structures are in virial equilibrium? Due to the lack of velocity information, $M-R$ relations cannot be directly linked to the virial state. However, having the linewidth - size relation of $\sigma_v\propto R^{0.5}$ satisfied (\citealt{Larson1981,Myers1983,Solomon1987,Falgarone2009}), the $M\propto R^2$ relation is suggested to imply that the cloud structures are in virial equilibrium (\citealt{Larson1981,Solomon1987}). However, as we discussed above, the $M\propto R^2$ relation does not necessarily mean a density profile of $\rho \propto R^{-1}$, and thus cannot be a straightforward indicator of virial equilibrium. When the $M \propto R^2$ relation is verified to imply $\rho \propto R^{-1}$, and if the structures also follow $\sigma_v\propto R^{0.5}$, the gravitational to kinetic energy ratio is a constant, and means virial equilibrium if that constant is about 2 (\citealt{Myers1988,Ballesteros-Paredes2006}). %Whether the structures are in virial equilibrium depends on the scaling coefficients of both the $M-R$ and the $\sigma_v-R$ relation and cannot be inferred only from $M\propto R^2$.

\section{Summary}\label{sec:conclu}

Using the column density map from \cite{Cao2019}, we obtain $N(R)$ profiles of 135 dense structures in Cygnus-X. At $0.1-10\space\rm pc$, all the structures have broken power-law $N(R)$ profiles, suggesting their dense core + diffuse cloud nature. With the transition at approximately $0.8\space\rm pc$ and $10^{22}\rm \space cm^{-2}$, the $N(R)$ profiles have a power-law index of $0.63 \pm 0.59$ at small radii, and $0.11 \pm 0.20$ at large radii.

We explore the $M-R$ relation using the broken power-law $N(R)$ profiles. Both the $N(R)$ profiles and the column density threshold $N_0$ determine the shape of the $M-R$ relation: for $N_0 > N_{\rm TP}$, we find (1) constant $N(R)$ power-law index and $N_0$ lead to $M\propto R^2$, (2) the $N(R)$ index with a wide range steepens the $M-R$ relation, (3) $N_{\rm TP}$ and $R_{\rm TP}$ with wider ranges make the data points in the $M-R$ plot spread out along loci following $M\propto R^2$, (4) $N_0$ with a wide range tend to make the $M-R$ relation follow $M(R)$ profiles. For $N_0<N_{\rm TP}$, the fact that $n$ is larger than $m$ makes the $M-R$ index slightly lower than 2. 
We apply $N_0$ with different means and ranges to the 135 Cygnus-X clumps and obtain $M-R$ relations with power-law indexes ranging from $1.4$ to $2.4$. 

From the observational perspective, the column density threshold $N_0$ in extracting cloud structures plays a crucial role in shaping the $M-R$ relation. With a constant $N_0$, the $M-R$ relation cannot be a probe of the density profile. Its $M-R$ index can be slightly less than 2 (when $N_0<N_{\rm TP}$), equal to 2 (when $N_0\ge N_{\rm TP}$ and $n$ has a tight distribution), and larger than 2 (when $N_0\ge N_{\rm TP}$ and $n$ has a wide range). For the cases with $N_0$ having a wide distribution and the data were not significantly affected by line-of-sight contamination or limited by small dynamical ranges, the $M-R$ relation can to large extent be a manifestation of the density profile. 

\begin{acknowledgements}
This work was supported by National Key R\&D Program of China No. 2017YFA0402600. We acknowledge the support from National Natural Science Foundation of China (NSFC) through grants U1731237, 11473011, 11590781 and 11629302.
\end{acknowledgements}

\bibliographystyle{raa}
\bibliography{2022-0092}{}

\label{lastpage}

\end{document}